\documentstyle[preprint,aps,prd]{revtex}

\begin{document}
\preprint{WU~B~94-13}
\draft
\title{  Proton-proton elastic scattering;\\
Landshoff contributions in the diquark model}
\author{ Rainer Jakob}
\address{Fachbereich Physik, Universit\"at Wuppertal,\\
D-42097 Wuppertal, Germany\footnote
{Supported by the Deutsche Forschungsgemeinschaft}}
\date{\today}
\maketitle

\begin{abstract}
Independent multiple scattering (`Landshoff') contributions
to proton-proton elastic scattering at wide angles are calculated
in the quark-diquark model. Results confirm previous observations
about the magnitude of these contributions. The use of the
quark-diquark model extends the applicability of perturbative QCD
calculations down to lower values of momentum transfer
substantially.
\end{abstract}


\section{Introduction}

Despite tiny cross sections and corresponding
difficulties for experimentalists, exclusive processes are the
natural approach to study the composite character of  hadrons. For
large momentum transfer the
wave functions deeply penetrate each other without
producing a torrent of secondary particles in the final
state. Thus compositeness is probed without destroying
the observed configuration.

Perturbative Quantum Chromodynamics
(PQCD) in the framework of the hard scattering picture (HSP)
\cite{Bro:75,Lep:80,Efr:80} is
the generally accepted theory to describe exclusive
processes at large momentum transfer. Factorization
of long- and short-range physics, the basic assumption of the
HSP, is reflected in the fact that exclusive
quantities are expressed as convolutions of process
independent distribution amplitudes (DA) with a
perturbative, hard amplitude for the scattering
of nearly collinear constituents.

The applicability of PQCD at intermediate
momentum transfer of a few GeV, where experimental
data are available, is
a matter of passionate controversy \cite{Isg:84,Dit:81}. The
overall momentum transfer in the process has
to be shared between the constituents in order
to align them suitably for
subsequent hadronization into the final state. Consequently, the
corresponding strong coupling in parts of the process may
become too large for reasonable use of
perturbative methods. In particular, this is the case,
when the momentum of a hadron is unequally shared between
it's constituents.

Hard elastic proton-proton scattering will certainly be a
cornerstone of investigations about the
hadronic structure. Unfortunately, the
relative complexity of the scattering of composite objects off
each other --- even if one only takes into account the valence
quark Fock states and diagrams on the Born-level --- has
prevented the complete cross sections from
being calculated up to now. The complexity is revealed in the
huge number of diagrams to be calculated ($\simeq$ in
the order of 100,000), as
well as in the occurence of pinch-singularities, which are
closely related to the existence of independent, multiple
scattering (`Landshoff' \cite{Lan:74}) contributions.

In a novel treatment of the `Landshoff' mechanism
in elastic proton-proton scattering Botts and
Sterman \cite{Bot:89,Bot:91} pointed out
the need for taking into account transverse momenta
in the HSP, which have been neglected before. The role of
transverse momenta in the `Landshoff' mechanism
is manifold: The energy dependence of the cross section is
understandable when one takes into consideration the
scaling behavior of momentum components transverse to
the scattering plane. Also, as has been shown in \cite{Bot:89}, the
the way transverse momenta are dealt with is decisive in
edriving a factorized formula for the scattering
amplitude. Soft, gluonic (`Sudakov' \cite{Sud:56}) corrections have
been resummed by the use of renormalization group techniques.
Here, the transverse separation between constituents (i.e. the
conjugate variable of the transverse momentum) acts as
an infrared cut-off and provides finiteness of the
results of loop-integrations. The resulting `Sudakov' factor
leads to a suppression of the scattering amplitude
and, thus, affects
the probability for a proton to contribute to elastic
scattering, depending on the transverse separation
between the constituents of the proton.

The work of Botts and Sterman \cite{Bot:89} about the
`Landshoff' mechanism
initiated an approach \cite{LiS:92,Li:92}, which
is addressed to refute the above mentioned
criticism \cite{Isg:84,Dit:81} of the applicability of
perturbative methods by modifying and improving on
the HSP. The basic idea, which has been demonstrated
for the calculation of electromagnetic form
factors in \cite{LiS:92,Li:92}, is to take into account
the transverse momenta
consistently also in the hard
scattering amplitude, where it still had been neglected in
\cite{Bot:89,Bot:91}. Dangerous, soft integration regions, where
the validity of perturbative formulas becomes doubtful, are
damped by the `Sudakov' corrections. Thus, self-consistency
of the perturbative calculation is achieved in the modified HSP
even for momentum transfers as low as a few
GeV. Here, self-consistency is meant in the sense that the
bulk of the results is derived with reasonable small values
for the strong coupling. Additionally,
the non-perturbative, intrinsic transverse
structure turns out to be important, as it strengthens the
suppression of soft regions. On the other hand, it provides a
substantially smaller pertubative result
\cite{Jak:93a,Bol:94}.

In the course of these developments the role of transverse
momenta in hard scattering processes, and correspondingly
the transverse structure of hadrons, has
received a lot of attention
\cite{Bot:89,Bot:91,LiS:92,Li:92,Hye:92,LiC:93,Zhi:93,Gou:94}.

In the present paper the `Landshoff' contributions of
proton-proton elastic scattering at wide angles are calculated
within a model in which the proton is
considered as a quark-diquark system. The treatment of two
correlated quarks as an effektive diquark is a possibility to
cope with non-perturbative effects still present in the
kinematic range of interest.
A systematic study of photon-proton reactions has been carried out
in the quark-diquark model: form factors in the space-like
and time-like region, real and virtual compton scattering,
two-photon-annihilations into proton-antiproton as well as
photoproduction of mesons \cite{Kro:88,Kro:93,Jak:93b}.

The motivation for the present investigation in the quark-diquark
model is the hope for an improvement of the applicability of the HSP
down to lower energies, compared to observations
made by Botts \cite{Bot:91} in the pure
quark picture. On the other hand, the reduction of complexity
(two constituents instead of three to deal with in the valence Fock
states) results in an technical simplification, which is not dramatically
for the `Landshoff' contributions, but might become decisive for
a future attempt to calculate all HSP diagrams (several 100's of
diagrams instead of several 100,000's).

The paper is organized as follows. In Sec. II the elements
of the quark-diquark model necessary for the present calculation
are briefly introduced. The mechanism of
independent scatterings is envisaged in Sec. III, gluonic
`Sudakov' corrections in the context of the quark-diquark
model are discussed and a factorized formula for helicity
amplitudes for the `Landshoff' contribution to elastic
proton-proton scattering at wide angles is given. Sec. IV contains
a discussion of numerical results. Conclusions
are given in Sec. V.

\section{The quark-diquark model}

The basic assumption of the diquark model is the clustering of
two of the three valence quarks in a baryon on an intermediate
energy scale, which allows to describe these two quarks, including
correlation effects, as an effective particle, the
diquark. Hence, some non-perturbative effects
still present on this intermediate scale are taken into
account. The coupling of spin-$1/2$ and flavor (isospin-$1/2$) wave
functions of two quarks leads to scalar and vector diquark
wave functions. The symmetry of proton wave functions
requires the spin and
the flavor parts of the diquark wave functions
to have the same symmetry. In this paper
only the scalar sector of the model is
considered, which is known to give rise to the bulk of numerical
results. The vector sector is esssential for spin effects, but
may be negligible for cross section results. The
quark-scalar diquark (S) Fock state contribution to the
proton state as a function of the usual longitudinal momentum
fraction $x$ and transverse momentum $\vec k_\perp$ of the
quark with respect to the proton's momentum $P$ is
\begin{equation}
\label{eq:states}
|P,\lambda\rangle_{Sq}=
\int\frac{dx\,d^{\;\!2}k_\perp}{16\pi^3\,\sqrt{x\,x'}}\;
\Psi_S(x,\vec k_\perp)\;
| S(x',\vec k_{\perp}') \;
  u_\lambda (x,\vec k_\perp) \rangle
\quad ,
\end{equation}
where $x'=1-x$ is the longitudinal momentum fraction of the diquark
and $\vec k_{\perp}'$ it's transverse momentum.

The dynamical content of the model is invoked by treating
the diquark as an elementary particle with corresponding
Feynman rules for the propagator of a scalar diquark and
the gluon diquark vertex respectively as
\begin{equation}
propagator:\quad
\frac{i}{p^2-{m_S}^2+i\varepsilon}\delta_{ij}
\qquad\qquad
SgS-vertex:\quad
i\,g_s\,t^a\:(p_1+p_2)_\mu
\qquad ,
\end{equation}
where $g_s=\sqrt{4\pi\alpha_s}$ is the QCD coupling, $i,j$ are
color indices and $t^a=\lambda^a/2$ are the Gell-Mann color
matrices. Diquarks are in an antitriplett color state, as is
necessary to form a color neutral baryon out of a diquark and
a single colored quark. The
composite nature of the diquarks is taken into account by
the introduction of phenomenological vertex functions which
may be parametrized as
\begin{equation}
F_S (Q^2) = \delta_S \; \left( \frac{{Q_S}^2}{{Q_S}^2+Q^2} \right)
\qquad
\delta_S = \left\{ \begin{array}{c c c}
1 & \quad\mbox{for}\quad & Q^2  <  {Q_S}^2 \\
\alpha_s(Q^2)/\alpha_s({Q_S}^2) & \quad\mbox{for}\quad & Q^2 \ge {Q_S}^2
\end{array}
\right. \quad,
\end{equation}
where $Q^2$ is the modulus of the squared momentum of the
gluon entering the vertex.
This form is choosen to ensure the diquark model to evolve into the
pure quark HSP in the limit $Q^2\to\infty$.

\section{`Landshoff' contributions to \lowercase{pp}
          elastic scattering}

\subsection{The `Landshoff' mechanism}

Independent scatterings in exclusive processes occur when
pairs of constituents accidently scatter by the same angle. In
this case the momentum transfer has not to be distributed
in the hadrons any further in order to guarantee nearly collinear
outgoing constituents, which are able to hadronize again; the
outgoing constituents are already suitably aligned by chance in
this special kinematical situation. This
results in a lower minimal number of gluons to be exchanged as
compared to the minimal number of gluons necessary in a general
HSP diagram. Consequently, with increasing energy `Landshoff'
contributions do not decrease according to `dimensional counting'
rules \cite{Bro:73}, but a bit slower.

This, so-called, `Landshoff effect' may be explained by the
observation that components of momenta transverse to the
scattering plane exhibit a different scale dependence as
compared to the other components. The reason for
this behavior is that independent scatterings
can be spatially separated in the direction transverse to
the scattering plane. On the contrary, with respect to directions
in the scattering plane the independent hard scatterings are
restricted to take place in a small region, the extension of which
is invers proportional to the center of mass
energy, i.e. proportional to $1/Q$.

The scaling behavior may be illustrated by considering the kinematics
of a `Landshoff' process.
In the quark-diquark model two types of diagrams
contribute as indicated in Fig.~\ref{fig1}. Fig.~\ref{fig1}a
correspond to a independent quark-quark and a diquark-diquark
scattering, whereas Fig.~\ref{fig1}b shows two
independent quark-diquark scatterings. The kinematics
of Fig.~\ref{fig1}a will be discussed in the following, kinematics
of Fig.~\ref{fig1}b can be infered from the former
by substitutions. Neglecting
masses and choosing the scattering plane to be
the ($z$-$x$)-plane the momenta in the hadronic center of mass system
are given as
\begin{eqnarray}
\label{eq:hadrmom}
P_1&=&(Q,0,0,Q) \hspace{26mm}
P_2=(Q,0,0,-Q) \nonumber\\
P_3&=&(Q,Q\sin\theta,0,Q\cos\theta) \qquad
P_4=(Q,-Q\sin\theta,0,-Q\cos\theta)
\quad .
\end{eqnarray}
The internal quark momenta (see Fig.~\ref{fig1}a) may
be parametrized as
\begin{eqnarray}
\label{eq:partmom}
\vec p_1=x_1 \vec P_1+\vec k_1
\quad &\Rightarrow & \;
p_1=(x_1Q+\sigma_1/Q~,\hspace{13.3mm}k_{1x}\hspace{13.3mm},
\nonumber\\
\vec p_2=x_2 \vec P_2+\vec k_2
\quad &\Rightarrow &\;
p_2=(x_2Q+\sigma_2/Q~,\hspace{13.3mm}k_{2x}\hspace{13.3mm},
\nonumber\\
\vec p_3=x_3 \vec P_3+\vec k_3
\quad &\Rightarrow &\;
p_3=(x_3Q+\sigma_3/Q~,~~~x_3Q\sin\theta+k_{3x}~,
\nonumber\\
\vec p_4=x_4 \vec P_4+\vec k_4
\quad &\Rightarrow &\;
p_4=(x_4Q+\sigma_4/Q~,~-x_4Q\sin\theta+k_{4x}~,~k_{4y}~,~
-x_4Q\cos\theta+k_{4z})
\; .
\end{eqnarray}
Internal momenta of the diquarks, ${p_i'}$, have an
analogous form. In the energy components extra
terms, $\sigma_i/Q$, are included to allow for
`on-shellness' of the quarks. Assuming
all transverse momenta (and, therefore, all $\sigma_i$
induced in the energy components by the transverse momenta) to
be small compared to Q, the four momentum conservation for
the quark-quark scattering reads
\begin{equation}
\label{eq:delta}
\delta^{(4)}(p_1+p_2-p_3-p_4)
\simeq \frac{
\delta\left(x_1-x_3\right)\delta\left(x_2-x_4\right)
\delta\left(x_1-x_2\right)}{2Q^3\sin\theta}
\delta\left(k_{1y}+k_{2y}-k_{3y}-k_{4y}\right)
\quad .
\end{equation}

All longitudinal momentum fractions, $x_i$, involved
in the quark-quark scattering are constrained to be equal. This
is characteristic for the special kinematical
situation in the `Landshoff' mechanism. The usual, hadronic
Mandelstam variables take the values $s=4Q^2$ and
$t=-2Q^2(1-\cos\theta)$. Their
partonic equivalents are approximated by
$\hat s\simeq x^2s ; \hat t\simeq x^2t ; \hat s'\simeq x'^2s ;
\hat t'\simeq x'^2t$. Hence, both partonic scattering angles
are equal to the scattering angle of the hadronic process
and, therefore, aligned constituents keep aligned during the
scattering process. The power of $Q$ in the denominator
of Eq.~(\ref{eq:delta}) is determined by the scaling behavior of
the energy component and the two momentum components
in the scattering plane. The
momentum conservation transverse to the scattering plane
(here: in the $y$-direction) is independent of the $Q$-scale, as
can be seen from Eq.~(\ref{eq:delta}).

The energy dependence of the hadronic scattering amplitude for the
`Landshoff' process may now be summed up as
\begin{equation}
M_{fi}\sim Q^{-3}\;F^2(Q^2)
\quad \mbox{(modulo logs)}
\quad .
\end{equation}
The factor $Q^{-3}$ originates from the momentum conservation
for the quark-quark scattering, Eq.~(\ref{eq:delta}). The momentum
conservation for the diquark-diquark scattering has not to be
considered separately, because the overall (hadronic) momentum
conservation automatically implies it, if the one for the quark-quark
scattering holds. The second factor, $F^2(Q^2)$, stems from
the diquark-diquark scattering, whereas the quark-quark scattering
only depends on the scattering angle. The hadronic wave functions
depend only logarithmically on the energy
scale. Consequently, the `Landshoff' contributions to the
differential cross section for elastic proton-proton scattering
in the quark-diquark description behave as
\begin{equation}
\label{scalingA}
\frac{d\sigma}{dt}=f(s/t)\,\cdot\,s^{-5}\,\cdot\,{F_S}^4(Q^2)
\qquad
\longrightarrow
f(s/t)\,\cdot\,s^{-9}
\qquad \mbox{for}\quad s\to\infty
\quad .
\end{equation}

This has to be compared with the predictions from the
`dimensional counting' rules:
Inserting one additional hard gluon in a `Landshoff' diagram
converts it into a HSP diagram. Then, there are no longer two
separate momentum conservations, the $Q^{-3}$-dependence
caused by the $\delta$-function is dropped. But instead, the
additional elements of the Feynman diagram give rise to
a factor $Q^{-4}$. (The insertion of a virtual gluon between two
quark lines, for example, introduces two quark propagators, two
quark-gluon vertices, and the gluon propagator, which
asymptotically scale like $\{Q^{-1}\}^2$, $\{Q^{\,0}\}^2$,
and $\{Q^{-2}\}^0$, respectively.) Thus, the
amplitude for a general HSP diagram behaves as
$Q^{-4}\,{F_S}^2(Q^2)$, which leads to an $s^{-10}$-behavior
for the differential cross section asymptotically.

\subsection{`Sudakov' corrections}

The leading radiative corrections to elastic proton-proton
scattering are similar in form to vertex loop-corrections.
For the case of QED Sudakov \cite{Sud:56} has shown that the coincidence
of `soft' and `collinear' divergencies in vertex corrections
typically leads to double logarithmic terms. Infrared divergencies
are regulated in these calculations by allowing for small
virtualities of external fermion lines. A similar form
has been derived for QCD vertex corrections \cite{Cor:76}, where
the non-abelian character of QCD is reflected in the
appearance of $\ln(\ln(q^2/m^2))$ terms. Higher order
of loop-corrections may be taken into account by exponentiating
single loop results \cite{Bel:80}.

In the quark-diquark picture `Sudakov' corrections to proton-proton
scattering are very similar to the corrections in the pion-pion
case, due to the fact, that diquarks carry the same color as
antiquarks. In Fig.~\ref{fig2} two types of gluonic corrections
are indicated. In axial gauge leading logarithms are given by
corrections of type I, which may be factorized into the wave
functions. Corrections of type II, which are non-factorizable, result
in non-leading logarithms.

A single-loop calculation in leading logarithm approximation
has been carried out for gluonic corrections to the
proton wave function in the quark-diquark
model. Exponentiating the result to account
for higher loops (but not
for non-leading logarithms) leads to a supression factor
\begin{equation}
\label{eq:sudfac}
\exp\left[-S(x,b,Q)\right]=
\exp\left[-s(x,b,Q)-s(1-x,b,Q)\right]
\end{equation}
with
\begin{equation}
\label{eq:sudfct}
s(x,b,Q)=\frac{C_F}{2\beta_1}\left\{
\ln\left(\frac{x\sqrt{2}Q}{\Lambda_{QCD}}\right)
\ln\left[\frac{\ln\left(x\sqrt{2}Q/\Lambda_{QCD}\right)}
{-\ln\left(b\Lambda_{QCD}\right)}\right]
-\ln\left(\frac{x\sqrt{2}Q}{\Lambda_{QCD}}\right)
-\ln\left(b\Lambda_{QCD})\right)\right\}
\quad ,
\end{equation}
where $C_F=4/3$ is the color factor and
$\beta_1=(11-2/3 n_f)/4$. Throughout
this paper $n_f=3$ and $\Lambda_{QCD}=200\,$MeV is used.
The result in Eq.~(\ref{eq:sudfac}) and Eq.~(\ref{eq:sudfct})
is equal to the correction for a pion wave function and, hence,
confirms that `Sudakov' corrections depend on color and not
on spin.

An essential point in Eq.~(\ref{eq:sudfct}) is the appearance of
the `impact parameter' $b$, which acts as an infrared cut-off. The
physical intuitive picture is
that the proton is viewed as a color dipole, formed by a quark
and a diquark. Therefore, the momentum range of soft
gluons, contributing to the corrections, is limited: The
upper limit is given by the large component of the quark
(diquark) momentum, i.e.
$x\sqrt{2}Q$ or $x'\sqrt{2}Q$, respectively. Harder gluons are
considered as higher order corrections to the hard scatterings and
not as a part of the soft `Sudakov' corrections. The lower limit
is induced by the inverse of the transverse separation of the
color charges, $1/b$. Gluons with wave lengths larger than
the dipole parameter $b$ effectively `see' a color
neutral object and decouple from the proton.
The larger the range of momenta between these two limits for a given
configuration, the stronger is the suppression by the
`Sudakov' factor. For a very small transverse separation the
infrared limit $1/b$ is close to the upper limit; there is
no suppression. A larger value of $b$ results in
a strong suppression. In Fig.~\ref{fig3} the
`Sudakov' factor, Eq.~(\ref{eq:sudfac}) is
displayed for a given value of $x=0.5$ and different
values of $Q$. Clearly the suppression tends to force
$b$ to zero for increasing $Q$.

Although the tendency of the `Sudakov' corrections
to keep colored constituents together is somehow similar
to the effect of confinement, it should be emphasized, that
Eq.~(\ref{eq:sudfac}) and Eq.~(\ref{eq:sudfct}) are entirely
perturbative. The `Sudakov' factor describes
the fact, that the probability for
a scattering process to take place in the exclusive channel is
decreasing with increasing spatial separation. It should not be
mixed up with the non-perturbative effect of confinement.

Resummation techniques based on renormalization
group equations have been developed to take into account leading
as well as non-leading logarithms to all orders
\cite{Col:80,Sen:81,Bot:89}. Working in a phenomenological
model like the quark-diquark model, it seems
reasonable to take a pragmatic point of view: Only the
exponentiated, leading-logarithmic corrections of
Eq.~(\ref{eq:sudfac}) and Eq.~(\ref{eq:sudfct})
are considered in the present calculations. Tacitly
it is assumed
that the neglection of non-leading corrections is an
acceptable approximation, as is indicated by the
results of resummations in the pure quark picture.

Mueller \cite{Mue:81} and Botts and Sterman \cite{Bot:89}
have shown, for the cases of pion-pion
and proton-proton scattering in the pure quark
picture, that gluonic `Sudakov' corrections to the
`Landshoff' contributions shift the power of the asymptotic
behavior near to the `dimensional counting' expectation.
Their arguments can readily be transfered to the present case
of proton-proton scattering viewed in the diquark model:
Assuming, for the moment, that the only $b$-dependence of
the amplitude is contained in the `Sudakov' factors
Eq.~(\ref{eq:sudfac}) of the
four proton wave functions, the integration over the
$b$-space can be estimated by insertion of
Eq.~(\ref{eq:sudfct}) and the use of a saddle point approximation
in the form
\begin{equation}
\label{eq:saddle-point}
\int_0^\infty db \;
\exp\left[-4\, S(x,b,Q) \right]
\simeq
\frac{\sqrt{2\pi}}{\Lambda_{QCD}}\;
\frac{\sqrt{c}}{1+c}\;
\sqrt{\ln\left(\sqrt{2xx'}Q/\Lambda_{QCD}\right)}\;
\left(\frac{\sqrt{2xx'}Q}{\Lambda_{QCD}}\right)
^{-c\,\ln\,\left(\frac{1+c}{c}\right)}
\quad ,
\end{equation}
where
\begin{equation}
c=\frac{4\,C_F}{\beta_1}=
\frac{64}{27},
\qquad
b_{sp}=\frac{1}{\Lambda_{QCD}}
\left(\frac{\sqrt{2xx'}Q}{\Lambda_{QCD}}\right)
^{-\frac{c}{1+c}}
\quad .
\end{equation}
Thus, the
leading power of Q, induced in the amplitude
by the `Sudakov' corrections, is
given by $-c\,\ln\,\left(1+1/c\right)= -0.83$.
Consequently the power of $s$ in Eq.~(\ref{scalingA}) for
the differential cross section
is changed to $-9.83$, which will be not distinguishable
experimentally from a power $-10$ in the foreseeable future.

\subsection{Hadronic helicity amplitudes}

Using Eq.~(\ref{eq:states}) for the proton helicity states a
matrix element for the hadronic process reads
\begin{eqnarray}
\label{eq:Tmatrix}
\lefteqn{
\langle P_3P_4|T|P_2P_1\rangle=\int\prod_{i=1}^4\
\frac{dx_i\,d^{\;\!2}k_{\perp i}}{16\pi^3\,\sqrt{xx'}}\
\Psi^\ast_S(\vec p_3)\Psi^\ast_S(\vec p_4)
\Psi_S(\vec p_2)\Psi_S(\vec p_1)}
\nonumber\\
&&\hspace*{20mm}\times
\left\{
 \langle q_3q_4|\hat T|q_1q_2\rangle
 \langle S_3S_4|\hat T'|S_2S_1\rangle
+\langle q_3S_4|\hat T|q_2S_1\rangle
 \langle S_3q_4|\hat T'|S_2q_1\rangle\right\}
\end{eqnarray}
where $\hat T$ and $\hat T'$ denote the two partonic transition
matrices of the independent scatterings. $S_i$ and $q_i$
symbolize the $i$-th scalar diquark and the $i$-th quark, respectively.
The momenta of hadrons, $P_i$, quarks, $p_i$, and diquarks, $p'_i$,
are defined as indicated in Eq.~(\ref{eq:hadrmom})
and Eq.~(\ref{eq:partmom}). Reminding
the relation between transition matrix elements and Feynman
amplitudes
$T_{fi}=i\,(2\pi)^4\,\delta^{(4)}\left(\vec P_f-\vec P_i\right)\,M_{fi}$
and the fact that momentum conservation for each of the
independent partonic scatterings implies the overall
hadronic momentum conservation, Eq.~(\ref{eq:delta}) leads to
\begin{eqnarray}
\lefteqn{
\label{eq:Mfi}
M_{fi}(s,t)=
\frac{i}{(2\pi)^8}\,\frac{1}{2^5\,Q^3\,\sin\vartheta}
\int_0^1 \frac{dx}{x^2x'^2}
\int\prod_{i=1}^4d^{\;\!2}k_{\perp i}
\;\delta\left( k_{1y}+k_{2y}-k_{3y}-k_{4y}\right)
}\nonumber\\
&&\Big\{\Psi^\ast_S(x,\vec k_{\perp 4} )\Psi^\ast_S(x,\vec k_{\perp 3})
\;\hat M_{qq;qq}(x,\hat s,\hat t)
\hat M'_{SS;SS}\;(x,\hat s',\hat t')
\;\Psi_S(x,\vec k_{\perp 2})\Psi_S(x,\vec k_{\perp 1})
\nonumber\\
&&\;\; +
\Psi^\ast_S(x',-\vec k_{\perp 4} )\Psi^\ast_S(x,\vec k_{\perp 3})
\;\hat M_{qS;qS}(x,\hat s,\hat t)
\hat M'_{Sq;Sq}\;(x,\hat s',\hat t')
\;\Psi_S(x',-\vec k_{\perp 2} )\Psi_S(x,\vec k_{\perp 1})
\Big\}
\; ,
\end{eqnarray}
where $\hat M$ and $\hat M'$ denote the partonic amplitudes. Following
the basic idea of Botts and Sterman \cite{Bot:89} a
factorized formula can be derived, when the
remaining $\delta$-function in Eq.~(\ref{eq:Mfi}), which is caused
by the conservation of momentum components
transverse to the scattering plane, is expressed by it's
Fourier transform
\begin{equation}
\label{eq:deltaF}
\delta\left( k_{1y}+k_{2y}-k_{3y}-k_{4y}\right) =
\frac{1}{2\pi}\int_{-\infty}^\infty db
\; e^{i\,b\cdot\left( k_{1y}+k_{2y}-k_{3y}-k_{4y}\right)}
\quad .
\end{equation}
The terms $e^{i\,\cdot k_{iy}}$ may be reabsorbed together
with momentum integrationsver the $k_{iy}$ by the definition
of wave functions in the form
\begin{equation}
\label{eq:Psitilde}
\tilde\Psi(x,k_{ix},b)\equiv
\int \frac{d k_{iy}}{2\pi}\;
\Psi_S(x,\vec k_{\perp,i}) e^{-i\,b\cdot k_{iy}}
\quad .
\end{equation}
These wave functions $\tilde \Psi(x,k_{ix},b)$ are the Fourier
transforms of the old ones, $\Psi_S(x,\vec k_{\perp,i})$ with
respect to the $y$-component. Eq.~(\ref{eq:Psitilde})
defines the parameter $b$ to be the conjugate variable to the
transverse momentum $k_{iy}$. Hence, as was mentioned
above, $b$ may be associated with
the separation of quark and diquark in the proton in the
$y$-direction. The effects of soft gluonic `Sudakov' corrections
are taken into account in leading logarithm approximation at
this stage of the calculation by multiplying the wave
functions $\tilde \Psi(x,k_{ix},b)$ with the exponential
factor $\exp\left[S(x,b,Q)\right]$ of Eq.~(\ref{eq:sudfac}).

The insertion of Eq.~(\ref{eq:deltaF}) and the use of the
definition Eq.~(\ref{eq:Psitilde}) leads to the factorized
formula
\begin{eqnarray}
\lefteqn{
M_{fi}(s,t)=
\frac{i}{(2\pi)^5}\,\frac{1}{2^5\,Q^3\,\sin\theta}
\int_0^1 \frac{dx}{x^2x'^2}
\int db \int\sum_{i=1}^4 dk_{ix}
\;\;\exp\left[-4\,S(x,b,Q)\right]}
\nonumber\\
&&\times\Big\{
\tilde\Psi_S^\ast (x,k_{4x},b) \tilde\Psi_S^\ast (x,k_{3x},b)
\,\hat M_{qq;qq}(x,\hat s,\hat t)
  \hat M'_{SS;SS}(x,\hat s',\hat t')
\,\tilde\Psi_S(x,k_{2x},b) \tilde\Psi_S(x,k_{1x},b)
\nonumber\\
&&\;\;+
\tilde\Psi_S^\ast (x',k_{4x},b) \tilde\Psi_S^\ast (x,k_{3x},b)
\,\hat M_{qS;qS}(x,\hat s,\hat t)
  \hat M'_{Sq;Sq}(x,\hat s',\hat t')
\,\tilde\Psi_S(x',k_{2x},b) \tilde\Psi_S(x,k_{1x},b)
\Big\}
\end{eqnarray}
for the helicity amplitude. Note that it is the
inclusion of transverse momenta in the calculation which
provides the key to derive the factorized formula. This is
based on the simple fact that the Fourier transform of a
convolution integral factorizes. Furthermore, the
gluonic corrections are treated such that they are described by
exponential factors to the wave functions, which do not destroy
the factorization.\footnote{As was shown in \cite{Bot:89} the
factorized form even holds for loop corrections, which can't be
written as an exponential multiplying the
wave functions (type II. in Fig.~\ref{fig2}), when
a suitable `soft approximation' is used.}

To perform the integrations over transverse momenta
$k_{xi}$ and $k_{yi}$, the latter contained in the
definition of the $\tilde\Psi_S(x,k_{ix},b)$,
an ansatz for the wave functions has to be made. Here, the
choice
\begin {equation}
\label{eq:waveans}
\Psi_S(x,\vec k_\perp)=
f_S \; \phi_S(x)\; \Sigma (x,\vec k_\perp)
\end{equation}
is used where
\begin{equation}
\label{eq:transans}
\Sigma (x,\vec k_\perp)=16\pi^2 \,\beta^2\, g(x)
\;\exp\left[-g(x)\,\beta^2\,k_\perp^2\right]
\qquad\mbox{and}\quad
g(x)=1\quad\mbox{or}\quad 1/xx'
\quad.
\end{equation}
The transverse momentum dependence is modeled as a simple
Gaussian, where the case $g(x)=1$ assumes factorization of
longitudinal and transverse degrees of freedom and the case
$g(x)=1/(xx')$ is inspired by harmonic oscillator wave functions
transformed to the light-cone, which have been
conjectured to describe meson wave
functions~\cite{Lep:83}. Correspondingly
two DA's are used in the form
\begin{mathletters}
\label{eq:DAans}
\begin{equation}
\phi_A(x)=N_A\,xx'^3\,
\exp\left[-\beta^2\left(\frac{{m_q}^2}{x}
+\frac{{m_S}^2}{x'}\right)\right]
\qquad \mbox{for}\quad g(x)=1/xx'
\label{eq:DAansA}
\end{equation}
\begin{equation}
\phi_B(x)=N_B\,xx'^3
\hspace{56mm}\mbox{for}\quad g(x)=1
\quad .
\label{eq:DAansB}
\end{equation}
\end{mathletters}
The polynomial $\sim xx'^3$ is the equivalent to the asymptotic
DA $\sim x_1x_2x_3$ in the pure quark picture and is related
to the latter by integration over one degree of freedom.
The values for $N_A$ and $N_B$ are fixed by the normalization
condition $\int\phi(x)dx=1$.  The Gaussian ansatz for the
$k_\perp$-dependence models the unknown, intrinsic (non-perturbative)
transverse structure of the proton. Note that fixing the
oscillator parameter $\beta$ with a phenomenological input, like
the root mean square (r.m.s.) of the transverse
momentum, $\langle {k_\perp}^2 \rangle$, introduces a hadronization or
confinement scale.

With the
ansatz of Eq.~(\ref{eq:transans}) the transverse
momentum integrations lead to
\begin{eqnarray}
\label{eq:ampDA}
M_{fi}(s,t)&=&
\frac{i\,4\pi^3}{Q^3\,\sin\theta}
\;{f_S}^4\;\int_0^1 \frac{dx}{x^2x'^2}
\int db \;
\phi^4(x)\;\hat M(x,\hat s,\hat t)
\;\hat M'(x,\hat s',\hat t')
\nonumber\\
&&\hspace{15mm}\times
\exp\left[-\frac{b^2}{g(x)\beta^2}\right]
\,\cdot\,\exp\left[-4\,S(x,b,Q)\right]
\quad .
\end{eqnarray}
Eq.~(\ref{eq:ampDA}) displays the two exponential
suppression factors brought about by the intrinsic, non-perturbative
transverse structure and by the perturbative `Sudakov'
corrections. Obviously, the borderline between both
effects will not be clearcut in
nature. Nevertheless, Eq.~(\ref{eq:ampDA}) indicates the point
of view adopted in the present paper: The perturbative formula for
the `Sudakov' factor is taken literally even in regions where
perturbative calculations are known to become
invalid, i.e. $1/b$ as low as $\Lambda_{QCD}$. The intrinsic
transverse structure, represented by the Gaussian, gives a weight function
for the probability of finding transverse distances in a
proton. Configurations with large $b$ values, corresponding to the
soft regions mentioned above (i.e. $1/b\to\Lambda_{QCD}$), have
a tiny probability to be found. Hence, the
error induced in the calculation by retaining incorrectly the
perturbative `Sudakov' formula in the very soft region is expected to be
small.

It is instructive to take a closer look at the interplay of the both
exponentials in Eq.~(\ref{eq:ampDA}). In Fig.~\ref{fig3} the
Gaussian $\exp\left(-b^2/4\beta^2\right)$ (i.e. $g(x)=1$) for
the value $\beta^2=1.389\,{\rm GeV}^{-2}$ (see Sec. IV) and
the `Sudakov' factor $\exp\left[-S(x=0.5,b,Q)\right]$, the latter
for different values of $\ln(s/s_0)$
with $s_0\equiv 1\,\mbox{GeV}^2$, are shown for comparison. Clearly
for large values of $\ln(s/s_0)$ the `Sudakov' factor dominates
the product of both. Thus, the asymptotic behavior of the
cross-section as estimated by the saddle-point approximation
Eq.~(\ref{eq:saddle-point}), i.e. $d\sigma /dt \sim s^{-9.83}$ is not
affected by the additional intrinsic transverse
structure. However, in the region of $\ln(s/s_0)$ smaller than
$\simeq 5\,\mbox{GeV}^2$ the Gaussian dominates the product of
both exponentials. Hence, taking into account the intrinsic
transverse, structure, e.g. in the form of a $Q$-independent
Gaussian as in the present work, damps the $Q$-dependence infered
by the `Sudakov' factor at least in the region of presently
available data.

Using the Feynman rules of the quark-diquark model
the helicity amplitudes can be calculated. Only three of
them are non-zero and get contributions from 4 diagrams
of type Fig.~\ref{fig1}a and 2 diagrams of type
Fig.~\ref{fig1}b:
\begin{eqnarray}
M^{hadr.}_{\{\lambda\}}(s,t)&=&
\frac{i\,4^4\pi^5{f_S}^4}{9\,Q^3\,\sin\theta}
\int_0^1 \frac{dx}{x^2x'^2}\int_{-\infty}^\infty db
\;\exp\left[-\frac{b^2}{g(x)\beta^2}\right]
\;\exp\left[-4\,S(x,b,Q)\right]
\alpha_s(x^2t)\alpha_s(x'^2t)
\nonumber\\
&&\times
\left\{\phi^4(x)\,F_S^2(x'^2\, t)
P^{(i)}_{\{\lambda\}}(s,t)
-2\,\phi^2(x)\phi^2(x')\,F_S(x^2t)\,F_S(x'^2t)
P^{(ii)}_{\{\lambda\}}(s,t)\right\}
\; ,
\end{eqnarray}
where $\{\lambda\}$ denote the three sets of helicities
$(++,++)$, $(+-,+-)$, and $(-+,+-)$.
The expressions $P^{(i)}_{\{\Lambda\}}$ and
$P^{(ii)}_{\{\Lambda\}}$ are explicitly given as
\begin{eqnarray}
P^{(i)}_{++,++}(s,t)&=&
\left(\frac{s(s-u)}{t^2}+\frac{s(s-t)}{u^2}-\frac{s^2}{ut}\right)
\quad ; \quad
P^{(ii)}_{++,++}(s,t)=
\left(\frac{su}{t^2}-\frac{st}{u^2}\right)
\nonumber\\
P^{(i)}_{+-,+-}(s,t)&=&
\left(\frac{1}{3}\frac{u(s-t)}{ut}-\frac{u(s-u)}{t^2}\right)
\hspace{10.5mm}; \quad
P^{(ii)}_{+-,+-}(s,t)=
\left(\frac{su}{t^2}\right)
\nonumber\\
P^{(i)}_{-+,+-}(s,t)&=&
\left(\frac{t(s-t)}{u^2}-\frac{1}{3}\frac{t(s-u)}{ut}\right)
\hspace{12.5mm} ; \quad
P^{(ii)}_{-+,+-}(s,t)=
\left(\frac{st}{u^2}\right)
\quad .
\end{eqnarray}
Using these results the differential cross section
\begin{equation}
\left.\frac{d\sigma^{pp\to pp}}{dt}\right|_{Ldsh. \atop diquark}(s,t)=
\frac{1}{16\pi}\;\frac{1}{s(s-4{m_p}^2)}\;\frac{1}{4}
\left\{
\left| M_{++,++}\right|^2 + \left| M_{+-,+-}\right|^2
+ \left| M_{+-,-+}\right|^2
\right\}
\, .
\end{equation}
has been calulated for a  scattering angle of $90^o$.

\section{Numerical results}

Parameters for the quark-diquark wave functions are taken from
\cite{Jak:93b}: ${Q_S}=3.22\,\mbox{GeV}^2$ and
$\beta^2=0.247\,\mbox{GeV}^{-2}$ and $1.389\,\mbox{GeV}^{-2}$ for
wave functions (\ref{eq:DAansA}) and
(\ref{eq:DAansB}), respectively. These values for
the oscillator parameter correspond to a r.m.s. transverse
momentum, $\langle{k_\perp}^2\rangle^{1/2}$ of $600\,\mbox{MeV}$.
The value for $f_S=73.85\,$MeV has been fixed by fits to
the data of electromagnetic form factors of the
nucleons~\cite{Jak:93b}; masses
of $m_q=330\,$MeV and $m_S=580\,$MeV are used.

Soft end-point regions of integration over longitudinal momenta
($x\to 0\;\mbox{or}\;1$) with corresponding singularities in
the strong couplings and in the gluon propagators are
avoided by the introduction of a cut-off parameter $C$ and the condition
\begin{equation}
\label{eq:cutoff}
\xi\geq C \frac{\Lambda}{\sqrt{2}Q}
\qquad \mbox{for}\;\xi=x,x'
\quad .
\end{equation}
Independence from the cut-off serves as an indication for the
range of applicability of the formalism. In the region of
small $b$ values, $b\leq 1/\sqrt{2}xQ$, the Sudakov factor
$e^{-S(x,b,Q)}$ is set to unity, it's value at $b=1/\sqrt{2}xQ$.

Results for `Landshoff' conributions to the
differential cross sections at $90^o$ obtained with the
wave function from
Eqs.~(\ref{eq:waveans}), (\ref{eq:transans}), and
(\ref{eq:DAansA}) are
shown in Fig.~\ref{fig4}. The dimensionless quantity
$R(s)=d\sigma /dt|_{90^o}\times 10^{-8}
s^{10}s_0^{-8}$ is plotted,
which should become constant according to the `dimensional
counting rules'. Results obtained with wave function B are
very similar in shape and magnitude to the results
in Fig.~\ref{fig4}. They are smaller by some per cent and
the independence of the cut-off is shifted a bit to higher
$\ln(s/s_0)$. Since differencies are really tiny, a figure
by its own for it may be dispensed with.

The rise of the curves at lower values of $\ln(s/s_0)$ are
caused by the $Q$-dependence of the cut-prescription
and, more important, by the behavior
of the phenomenological vertex functions
$F_S(Q^2)$, which have not yet reached their
asymptotic $Q^{-2}$ behavior. The position of the
maximum is predominantly determined by the value of the diquark
parameter $Q_S$. The fallings of the curves are induced by
the behavior of the strong coupling; the cross section is
proportional to ${\alpha_s}^4$.

A comment about the experimental data should be made here:
The data reveal roughly the expected scaling with $s^{-10}$ modified
by, as it seems to be, an oscillation
about it. It has been speculated
\cite{Ani:92} that the data indeed do not show the beginning
of an oscillation, but a two peak structure, with the second peak
caused by diquark correlations in the proton. The present results
confirm the existence of a bump, even roughly peaked in
the energy region of interest. But the shape and magnitude of
this bump clearly disfavors this explanation for the
structure of the data.

A different explanation has been suggested some years
ago. Ralston and Pire \cite{Ral:82} concluded the existence of a
phase, proportional to $\ln(s/s_0)$, from
analyticity properties of gluonic corrections
and fitted the coefficients to the data. Botts and
Sterman \cite{Bot:89} showed
that corrections of type II in Fig.~\ref{fig2}, neglected
in the present paper, cause imaginary parts in the
`Sudakov' functions. However, the resulting phase is
proportional to the ratio $\ln(s/s_0)/\ln(b\Lambda)$, which
approaches a constant in the region where the saddle-point
approximation Eq.~(\ref{eq:saddle-point}) is valid, because
of the power law behavior of the
saddle-point $b_{sp}\sim s^{-0.35}$. Although the derivation of
the phase from corrections of type II in Fig.~\ref{fig2} is
beyond the scope of this paper, a comment can be made about
this problem: The inclusion of the intrinsic transverse
structure, neglected in previous papers, reconciles the
$s$-dependence of the phase in the region $\ln(s/s_0)\leq 5$.
Here, the Gaussian damps the `Sudakov' factor and the dominantly
contributing $b$-region is almost $s$-independent.

The magnitude of the results in Fig.~\ref{fig4} is roughly by a factor
of $10$ below the experimental data, but it is definitely not suppressed
by many order of magnitudes as has been presumed before
\cite{Lep:80}. In this sense the result is an independent
confirmation of the observations Botts made in the pure quark
picture \cite{Bot:91}. At $\ln(s/s_0)=5$ the results in \cite{Bot:91}
vary in the range of $R=0.07\ldots 9$ depending on the choosen
distribution amplitude and the value of the cut-off. This is
in accordance with the range of $R=0.5\ldots 1.5$ found in the
present calculation in the framework of the quark-diquark model.
The largest uncertainty in the magnitude of the results
is caused by the normalization of the wave functions. The
value $f_S=73.85\,$MeV is taken from fits to the
electromagnetic form factors of the nucleons \cite{Jak:93b}. These
have been done without assuming a concrete $k_\perp$-dependence, what
leads to a freedom to vary $f_S$ in a limited range. Assuming
a concrete $k_\perp$-dependence, like
the Gaussian in the present case, fixes
the relation between $f_S, \langle{k_\perp}^2\rangle$ and
$P_{qS}$, the probaility to find a proton as a system of a
quark and a scalar diquark. The presently used values of
$\langle{k_\perp}^2\rangle^{1/2}$ and $f_S$ correspond to
$P_{qS}\simeq 1$. Constraining, for example, this
probability to be $P_{qS}=0.5$
would change $f_S$ by a factor of $1/2$ and correspondingly
change the cross section by a factor of 1/16. These
considerations may indicate that the lacking knowledge about the
non-perturbative wave functions easily induces uncertainties
of one order of magnitude.

Comparison to data is, somehow, ambiguous because the results
show strong cut-off dependence in the region of the data.
Independence of the cut-off and therefore applicability of the
formalism is reached for $ln(s/s_0)\simeq 6 \;
(\mbox{i.e.}\;s\simeq 400\,\mbox{GeV}^2)$. This has to
be contrasted with a value of $ln(s/s_0)\simeq 8 \;
(\mbox{i.e.}\;s\simeq 3000\,\mbox{GeV}^2)$ given in \cite{Bot:91}.
Thus the hope of improving the applicability down to lower values
of $s$ by using the quark-diquark model is fulfilled.

A further improvement of applicability will surely be obtained
by taking into account also the transverse momenta in
the hard scattering amplitudes itself, which have been
neglected up to now. Thus,the strategy of \cite{LiS:92,Li:92},
developed for form factors, could be adopted; i.e. to
characterize soft regions by both, small $x$ and
small transverse momenta (or large $b$ values). These regions
are suppressed by the `Sudakov' factor. The transfer of this
concept to the present case of proton-proton scattering is
not straightforward, because the inclusion of transverse
momenta in the gluon propagators
would destroy the factorized form of Eq.(\ref{eq:ampDA}). What can
safely be done to improve the applicability of the calculation, is
the replacement of the arguments of the strong coupling
in the form
\begin{equation}
\label{eq:newarg}
\alpha_s\left(x^2t\right)\to
\alpha_s\left(\max\left(x^2t,1/b^2\right)\right)
\hspace{20mm}
\alpha_s\left(x'^2t\right)\to
\alpha_s\left(\max\left(x'^2t,1/b^2\right)\right)
\quad .
\end{equation}
Such, the largest scale in each independent hard process
determines the strength of the coupling. For
vanishingly small $x,x'$ the transverse scale takes over.

Results with these prescriptions, Eq.~(\ref{eq:newarg}) are
displayed in Fig.~\ref{fig5} in comparison with the
curves from Fig.~\ref{fig4}. Evidently they coincide
asymptotically. For small values of $\ln(s/s_0)$ the
modified version lies a bit below the results obtained with
a cut-off $C=1.1$. This effect is readily explained by noting
that the arguments of the coupling have become smaller on an
average. It's worth emphasizing that the modified result
(i.e. the solid line in Fig.~\ref{fig5}) is derived entirely
without a cut-off (or $C=0$). On the contrary, results
of calculations without
the replacement Eq.~(\ref{eq:newarg}) diverge for $C\leq1$. The
reliability of the modified calculation, and thus the effectivity
of `Sudakov' suppression of soft regions, can be checked by testing
the portion which has been obtained with reasonable small values of
the strong coupling, say $\alpha_s\leq 0.5$. It turns out
that at $\ln(s/s_0)=1.5$ 52\% and at $\ln(s/s_0)=3$ even 84\% of
the full result fulfill this criterion.

Clearly the replacement Eq.~(\ref{eq:newarg}) does not substitute
a calculation with all the transverse momentum dependence taken
into acount in the hard scatterings. But
results are quite encouraging that such a, still lacking, more
complete calculation will render the formalism reliable down to
the energy range of the data.

\section{Conclusions}

It has been emphasized that transverse momenta are the
key of understanding for multiple, independent scattering
processes with respect to factorization and their scaling
behavior. The present calculation of `Landshoff' contributions
to the elastic proton-proton scattering at wide angles
in the framework of the
quark-diquark model is an independent confirmation of
observations made before \cite{Bot:91} in the pure quark picture.
The magnitude of `Landshoff' contributions is small, but definitely
not suppressed by many orders of magnitude and, therefore, a priori
not negligible. The main uncertainties in the calculation stem from
our incomplete knowledge about non-perturbative wave-functions and
the applicability of the calculation, as indicated by cut-off
independence, is beyond the range of experimental access.
It has been shown that the phenomenological quark-diquark model,
as additional assumption to the hard scattering picture, improves
the range of applicability of the perturbative calculations
substantially down to smaller values
of $s$, but still outside the energy region of experimental data.

A way out of this problem by taking into account transverse
momenta scales, at least in the argument of the strong
coupling, has been discussed and numerically tested.
The results clearly indicate that an extension of the HSP
by taking into account transverse momenta consistently
(i.e. also in the hard scattering amplitudes) will lead to
self-consistency of the perturbative calculation
and, thus, will improve the reliability of the results down
to even very small values of $s$.

\acknowledgments
I would like to thank J.~Botts, T.~Gousset, B.~Pire and J.P.~Ralston
for clarifying
discussions and P.~Kroll for invaluable advice
and careful corrections.

%
%
%
\begin{figure}
\caption{Types of diagrams contributing to elastic proton-proton
scattering at wide angles via the `Landshoff' mechanism. Protons
are considered as quark-diquark systems; double
lines indicate the diquarks.}
\label{fig1}
\end{figure}
%
%
\begin{figure}
\caption{One-loop gluonic corrections to the diagrams of
Fig.~\protect\ref{fig1}. Type I. corrections may be factorized
into the wave functions. Type II. corrections are non-factorizable.}
\label{fig2}
\end{figure}
\begin{figure}
\caption{Sudakov factor of Eq.~(\protect\ref{eq:sudfac})
for $x=0.5$ and different values of $\ln(s/s_0)$ with
$s_0\equiv 1\,\mbox{GeV}^2$ (dashed and dashed-dotted lines). For
comparison the Gaussian
$\exp\left(-b^2/4\,\beta^2\right)$ caused by the intrinsic
transverse momentum dependence is also shown (solid line).}
\label{fig3}
\end{figure}
\begin{figure}
\caption{Elastic Proton-proton scattering at $90^o$. The
dimensionless quantity $R(s)=d\sigma /dt |_{90^o}\times
10^{-8}\,s^{10}\,s_0^{-8}$ obtained with wave function
(\protect\ref{eq:DAansA}) is plotted
against $\ln(s/s_0)$. Data
are taken from the data compilation \protect\cite{Bys:78}.}
\label{fig4}
\end{figure}
\begin{figure}
\caption{Comparison of modified
calculation, Eq.~(\protect\ref{eq:newarg}), with
the cut-off method (cf. Fig.~\protect\ref{fig4}). Both
calculations are done with wave
function (\protect\ref{eq:DAansA}).}
\label{fig5}
\end{figure}
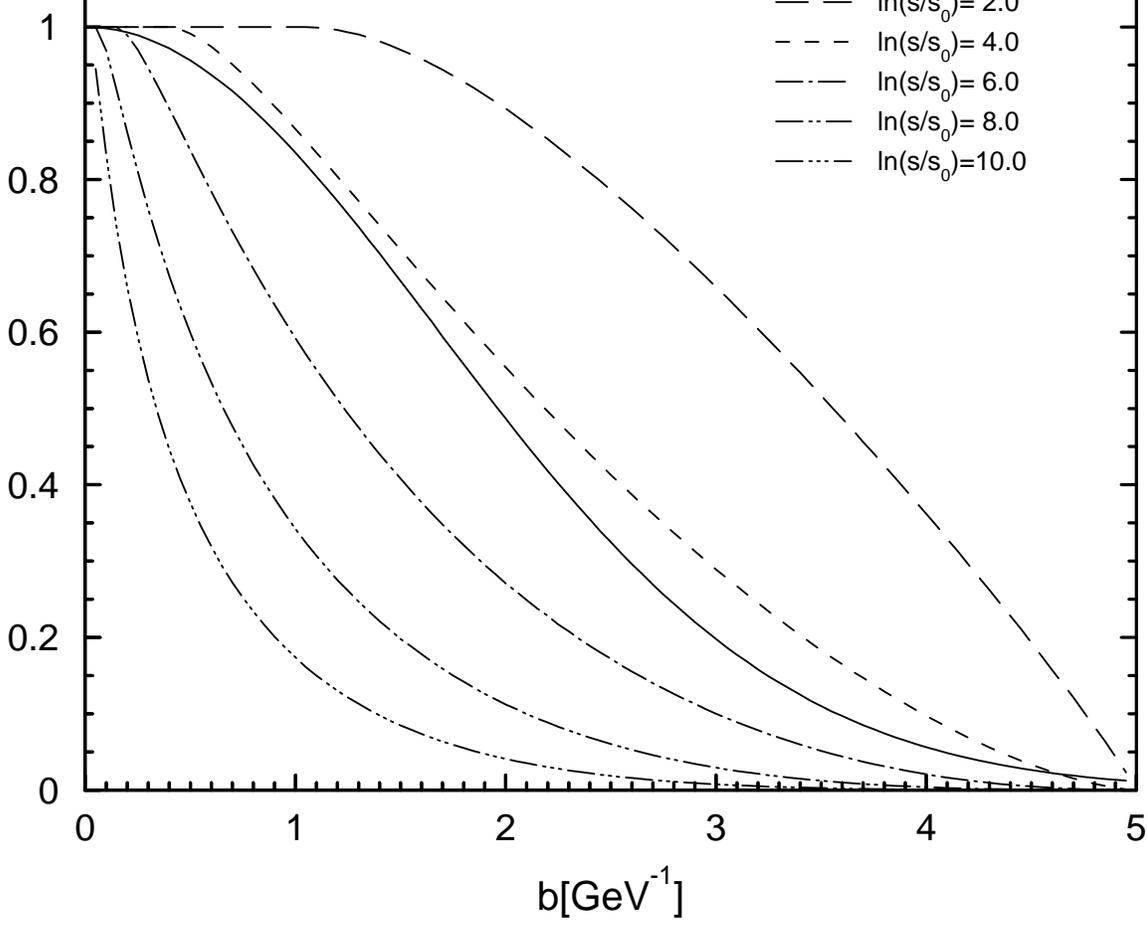
\end{document}